\documentclass[runningheads]{llncs}
\usepackage{graphicx}
\usepackage{amssymb}
\usepackage{blkarray}
\usepackage{amsmath}
\usepackage{comment} 
\begin{document}
\title{Tourists Profiling by Interest Analysis}
%
%
\author{Sonia Djebali\inst{1} \and Quentin Gabot\inst{1,2} \and Guillame Guérard\inst{1}
}
\authorrunning{S. Djebali et al.
}
\institute{Léonard De Vinci, Research Center, 92 916 Paris La Défense, France
\email{f\_name.l\_name@devinci.fr}
\and Research Student
\email{f\_name.l\_name@edu.devinci.fr}
}
\maketitle              
\begin{abstract}
With the recent digital revolution, analyzing of tourists' behaviors and research fields associated with it have changed profoundly. It is now easier to examine behaviors of tourists using digital traces they leave during their travels. The studies conducted on diverse aspects of tourism focus on quantitative aspects of digital traces to reach its conclusions. In this paper, we suggest a study focused on both qualitative and quantitative aspect of digital traces to understand the dynamics governing tourist behavior, especially those concerning attractions networks.

\end{abstract}
\section{Introduction}

Nowadays, tourism industry is considered as one of the largest and fastest growing industries \cite{cooper2007contemporary}. In~2019, \textit{World Tourism Organisation} UNWTO\footnote{International tourism Growth continues outpacing the global economy: edition 2020.} has recorded $1.5$~billion international tourists, $4\%$~more than the previous year.
%

With the recent booming of digital tools and mobile internet technology, alternative sources of data to understand tourism behaviors and to capture tourists' experiences have emerged. Users of social networks, 
tend to share openly and frequently photos, videos, reviews and recommendations of their travels and their experiences. Thus, when users share photos or reviews, geographical information is included. These geo-located data represent tourism and sociological views~\cite{chareyron2009}. These social networks constitute an interesting observation field to analyze tourists' behavior evolution through time and space.

In social networks, graph models are widely used to represent an interaction between entities and their relations. 
A graph model allows to understand both local phenomena (i.e. node level) 
as well as global phenomena (i.e. graph level). 
One of an interesting way mostly used to analyze movement networks is to evaluate and detect community structure~\cite{palla2005uncovering}. 
Detecting communities in networks provides a means of coarse-graining interactions between entities and offers a more interpretable summary of a complex network.

%
In literature, studies carried out on social networks use a quantitative dimension of digital traces as a research hypothesis \cite{kaufmann2019analysis}. 
Models developed focus on the question of the number of tourists' 
without considering a qualitative analysis and the interest of users to make a given  action.
Interest analysis allows us to identify trends that are not perceptible with quantitative analysis. Being able to define another kind of relationship between two places visited in a city will not give the same result as observing  flows of tourists between these two places.
%

To better understand tourists movements and their interest to consider this movements, we propose in this paper to detect tourist communities to perform tourist profiling based not on the quantitative aspect but also on the qualitative aspect i.e. the interest of the tourists to make movements or visits.
Our study takes advantage of existing quantitative methods to analyze how of tourists perform an action. Then, we enhance them with qualitative methods to analyze why tourists perform this action.
Thanks to a community detection and profiling, stakeholders will be able to better identify behaviors and thus target communities by proposing adapted excursions at the best places.

Our key contributions can be summarized as : \textbf{(1)} an automatic \texttt{Graph Movement} extraction methodology dedicated to graph of interest,
\textbf{(2)} the \texttt{Measure of Interest} using an interest measure as weighted in a graph. This measure is not dependent on  frequency or probability based on number of tourists, \textbf{(3)} \texttt{Spheres of influence} to compare a neighborhood of each visited place to determine a similarity matrix, \textbf{(4)} \texttt{Community detection} by a clustering on the similarity matrix provide profiles. A thorough study of tourists' profiles is compared to tourism management works.

In this paper, we will first relate in Section~\ref{sec:soa} comparable works on community detection network analysis. In Section~\ref{sec:graph} we formalize our graph data model. In  Section~\ref{sec:comm} we present our approach to perform a community tourist detection and tourist profiling.
Our model is implemented and is subject of a case study on a TripAdvisor dataset in Section \ref{sec:xp}. Finally, the Section \ref{Conclusion and future works} concludes this paper.

\section{State of the Art}
\label{sec:soa}

Social networks are a social structure where a set of social actors are connected by relationships. Due to the size of social networks with numerous and heterogeneous relationships, it has been becoming difficult to analyze them.
Therefore, several algorithms and graph theory concepts are used to study a structure of social networks \cite{knoke2019social}.
Most researches aiming at analyzing social networks enable to understand different social phenomena including social structures evolution, 
to measure an importance of nodes, to detect communities of nodes \cite{arenas2004community} that share some characteristics by looking metrics at the whole network cohesion \cite{kolaczyk2014statistical}, etc.

Social networks are paradigmatic examples of graphs with communities. The word community itself refers to a social context. 
The concept of community has no unified definition \cite{lancichinetti2009community}, 
most of the researchers have reached a consensus that communities in a network indicate groups of nodes, such that nodes within a group are connected more often than those across different groups \cite{xia2013survey}. 
Detecting community in networks is very hard and not yet satisfyingly solved, despite an huge effort of a large interdisciplinary community of scientists working on it over the past few years.

In the literature, several researches have been interested to detect of community in social networks to analyze and understand tourist behaviors.
To detect tourists community, social networks are modeled as a directed graph with a weight attached to each edge.
These values of weights can be representing, for example, frequencies of movement of tourists, probabilities value, rules metrics, etc. Based on this graph model, a detection community method and approach exposed below have been studied by many research fields.

\textit{Graph Partitioning}: This method consists to divide a graph into groups of predefined size, such that the number of links in a cluster is denser than the number of edges between the clusters \cite{Fortunato_2010}. The disadvantage of this method 
is to define the size of clusters in a relevant way.

\textit{Hierarchical Clustering.} 
Hierarchical clustering techniques are based on a nodes similarity measure \cite{friedman2001elements}. Theses techniques don't need a predefined size and number of communities. 
Hierarchical clustering techniques can be categorized into two classes. \textit{Agglomerative algorithms} starts by considering each node of a graph as a separate cluster and iteratively merge them based on high similarity and ends up with an unique community. And \textit{Divisive algorithms} start by the entire network as a distinct cluster and iteratively splits it by eliminating links joining nodes with low similarity and ends up with unique communities.

\textit{Modularity Optimisation Based.} Modality, is a quality function introduced by Newman and Girvan \cite{Nicosia2009ExtendingTD} to measure the quality of a partition of graph nodes. The larger the modularity value the better is the partition. The best-known methods within are: \textit{Greedy method of Newman} is an agglomerative method. Initially, each node belongs to a distinct module, then they are merged iteratively based on the modularity gain. And
\textit{Blondal’s Louvain algorithm} is an heuristic greedy algorithm for uncovering communities in complex weighted graphs \cite{clauset2004finding}. It is based on the modularity optimization.

\textit{Statistical inference-based methods.} These methods deduct properties of data\-sets, starting from a set of observations and model hypotheses. If the dataset is a graph, based on hypotheses on how nodes are connected to each other, the model has to fit the actual graph topology \cite{Fortunato_2010}. These methods attempt to find the best fit in a model to the graph, where the model assumes that nodes have some sort of classification, based on their connectivity patterns.
%

All the methods cited above are used to detect communities on network models as a directed weighted graph. Where  weights values represent frequencies, probabilities, labels, vectors, etc.
The probability weights is deduced from frequencies or is computed, thanks to data mining, by considering relation between nodes as rules. 
Those methods implement a measure like support, confidence or lift 
and return rules which satisfy some constraints.
To determine communities, one can analyze a graph as a Markov Chain \cite{baccar2019tourist} 
Most advances methods use adapted clustering or partitioning such as spectral clustering \cite{von2006tutorial}, label propagation \cite{raghavan2007near}.

In the tourism research field,  methods used for community detection are focused on detecting community based of the quantitative dimension i.e. frequencies movement of individuals or flows in a graph.
These methods form clusters based on clustering metrics (Silhouette, Dunn, etc.). However, the interpretation of clustering contents is unrepresentative of a community sharing a same interest or a profile but rather a mass of data.
In this study, we will focus on detection community to extract tourist profiling base on the interest of tourists to make a movement or visit by integrating the notion of overtourism \cite{capocchi2019overtourism}.

\section{Tourism Movement's Data Model}
\label{sec:graph}

To model tourism movement, we must consider visited locations, users information and their interactions. To study interactions on locations, we propose to model tourist data by a graph of movement.
Graphs rely on links between users and locations through their reviews. 

\paragraph{Data Types}

Our database is composed of users, reviews, and geo-located locations. A location is composed of longitude and latitude coordinates, type of location (hotel, restaurant, attraction) and a rating. Each location has been aligned with administrative areas (GADM)\footnote{GADM:\url{https://gadm.org/index.html}. $386,735$ administrative areas (country, region, department, district, city, and town).}.
A user is identified by nationality, age and is described by a timeline. A user timeline represents a chronological set of reviews from its first reviews to its last reviews. This timeline allows computing intermediate properties like time between two consecutive reviews i.e. consecutive visited places.
A review represents a note given by a user on a location at a given time.
\subsection{Sequences Dataset}
\label{dataset}

To study tourists’ movement of a given destination, we need to target tourists. For this, we focus only on users who visit at least one destination. Then we extract all their reviews to gather their circulation all over the world.
Tourists review several locations during their trip. A trip is a succession of days when a tourist writ at least one review per day, i.e. as soon as there is a day when a tourist does not write a review, its trip is considered broken.
However, a tourist may not review during a limited period during the same trip. The break can be canceled between two trips that took place in the similar country by a user. We consider a sequence is composed of reviews written at most at $7$~days apart \cite{gossling2018tourism}.
The method consists of merging two trips if they satisfy the following conditions: 
$\Delta{B}$ $\leqslant$ $\Delta{T_i}$  and  $\Delta{B}$ $\leqslant$ $\Delta{T_j}$ and $ L_{T_i} = F_{T_j} $.
Where $\Delta{B} \leq 7$ represents break's duration, $\Delta{T_i}$ and  $\Delta{T_j}$ present $i^{th}$ and $j^{th}$ trip's duration, $F_{T_j}$ represents the first country visited during $j^{th}$ trip and $L_{T_i}$ represents the last country visited during $i^{th}$ trip.
For each tourist, we build a set of trips based on its reviews. Each trip corresponds to sequence and presents a succession of locations 
in a temporal order of visit during a trip.

\subsection{Sequential Rule Mining}

Based on sequences dataset, we use a sequential rules mining algorithm to discover and to extract all existing rules in the dataset. The input structure is a set of sequences, and the output structure is the rules.
A sequential rule is a rule of the form $X \longrightarrow Y$ where $X$ and $Y$ are items. 
In our case a rule $X \longrightarrow Y$ is interpreted as, if location $X$ occurs then it will be \textit{directly followed} by the location $Y$.
To exploit the sequential rules, two measures are predominantly employed, \textit{support} and \textit{confidence}. The support of a rule $X \longrightarrow Y$ is how many sequences contains the location from $X$ followed directly by the location $Y$. The confidence of a rule $X \longrightarrow Y$ is the support of the rule divided by the number of sequences containing the location $X$.

To discover sequential rules appearing in sequence dataset, we use a \textit{TRuleGrowth} algorithm \cite{fournier2012mining} for its performance. The TRuleGrowth algorithm uses a pattern-growth approach for discovering sequential rules such that it can be much more efficient and scalable.
After, extracting all the rules from the sequences dataset, we use an \textit{Interest measure} to value interesting and useful rules for an effective decision making.

\subsection{Measure of interest}

To value interesting rules, there exist several measures developed in divers fields such as machine learning, social science, statistics, data mining, etc. Selecting a right measure for a given application is difficult because many measures may disagree with each other.
Each measure has several properties that make them unsuitable for other applications. In our case, the key properties one should consider selecting an accurate measure are:
\begin{enumerate}
\item The measure must distinguish the movement between location $X$ and $Y$ with that between location $Y$ and $X$. The measure must be \textbf{asymmetric}.

\item To understand the interest of tourists to perform a given movement, the measure must be \textbf{antisymmetry} to express the positive and the negative correlation resulting of the movement between two locations.

\item The measure must have a positive value if there is a positive correlation and equals zero if items are \textbf{statistically independent}.

\item Considering size of the dataset, the measure must be easy to implement, readable, editable and context-free with a \textbf{limit scale}. The measure must have an upper bound and a lower bound.

\item The quantity dimension 
mustn't be ignored. The measure will scale in a limited way with it. 
\end{enumerate}

The measure respects and corresponds to all these properties is \textit{Klosgen} \cite{klosgen1992problems}. 
The Klosgen measure is a normalization of the \textit{Added Value measure}, also called \textit{Pavillon Index} or \textit{Confidence}, \cite{sahar1999empirical}.
For a given rule $X \longrightarrow Y$ Added Value $ AV(X \Rightarrow Y)$ measure quantifies how much the probability of $Y$ increases when conditioning on the transactions that contain $X$ :
      $AV (X \Rightarrow Y) = confidence(X \Rightarrow Y) - support(Y))$. Where,
$confidence(X \Rightarrow Y) = \frac{|X \Rightarrow Y|}{|X|}$ and  $support(X)=|X|$. $|X|$ denotes the frequency of $X$ in the database.
The Added Value of a rule can be positive, negative or zero. The rules with a high Added Value are considering as interesting rules. Since the Added Value vary greatly depending of the support of the antecedent $X$ and the consequent $Y$, we normalize the Added Value measure by multiplying it by its squared root which gives Klosgen measure.
Then, we compute the Klosgen measure for each rule as follows:
$Klosgen(X \Rightarrow Y) =  \sqrt{supp(X \Rightarrow Y)} *AV(X \Rightarrow y)$

Using the Klosgen measure has two main advantages. The first is that,  the influence of support or confidence is lowered, which means that very low or very high support of the antecedent or the consequent have a low influence on the value of the measure (unlike lift or confidence) \cite{tan2004selecting}. The second is that the Klosgen measure is not a complex measure, which makes its meaning relatively clear, since it is presented as the normalization of the Added Value. The Added Value being itself part of the basic objective interest measures.

\subsection{Graph movement model}
\begin{figure}[t]
	\begin{minipage}[b]{0.45\linewidth}
		\centering
		\includegraphics[width=\columnwidth]{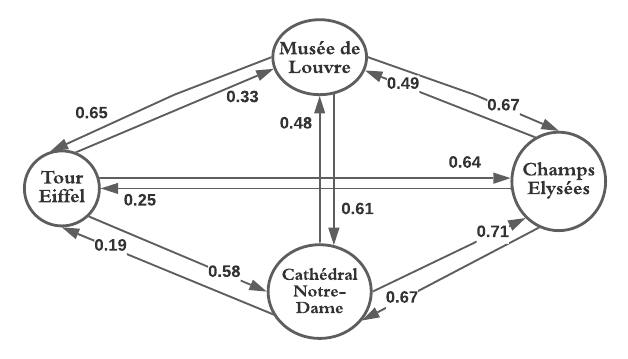}
		\caption{Directed weighted graph\\  G(V,E,W) with Klosgen
		values of\\ the four most visited 
		places of Paris.}
		\label{fig:basicgraph}
	\end{minipage}
	\hfill
	\begin{minipage}[b]{0.45\linewidth}
		\centering
		\includegraphics[width=\columnwidth]{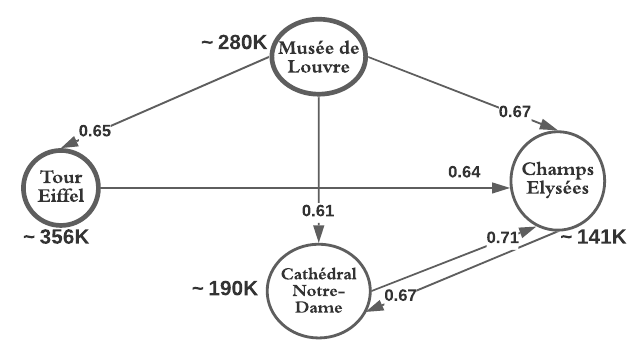}
		\caption{Weighted graph of the top $2$\\ 
		mainstream nodes with a Klosgen \\
		threshold at $0.6$.}
		\label{fig:graph2}
	\end{minipage}
\end{figure}

Based on the set of rules and their interest value calculating by Klosgen, we build a movement graph of tourists.
The movement graph $G( V, E, W)$ is a weighted directed graph. The nodes $V$ are a set of the antecedents and the consequent of the rules. These nodes present the locations visited by the tourists. The arcs $E$ are the association of the rules and present two direct consecutive locations. The weighted $W$ of the arcs is the values of the \textit{interest} measure of rules that represent interest of the tourists to move between two places i.e. two nodes.
The figure \ref{fig:basicgraph} shows an example of graph of movements of tourist fours most visited place in $Paris$.

\section{Community Detection}
\label{sec:comm}

Based of our graph model, to detect community, we first identify a \textbf{mainstream nodes}, the nodes having a most influential and strong attractiveness.
Secondly, we define a \textbf{neighborhood} of mainstream nodes. The neighbors of a node consist in a set of nodes that are connected to this node up to a \textit{certain distance}, i.e., the number of steps between a source node and its neighbors.
Then, for each mainstream node we build a \textbf{Sphere of Influence}. 
Finally, we define a \textbf{similarity measure} between spheres of influence, this measure is used to create \textbf{clusters} and then  detect communities. 

\subsection{Mainstream nodes}
We will define two kinds of nodes, \textit{mainstream nodes} having the most influential on the graph and \textit{secondary nodes} the other remaining nodes.
Based on graph of tourist movements $G(V,E,W)$, for each node of $V$ we calculate its support (i.e. number of occurrences) on the sequences dataset, described in section \ref{dataset}.
We select a subset of node $V'$ having a $K$ maximum value of support. The value of $k$ is fixed according to the number of nodes in the graph and the distribution of the support amount all the nodes.
We note $V''=V\setminus V'$ the secondary nodes as the rest of the nodes which are not mainstream.

Once mainstream nodes and secondary nodes are defined, to perform community detection, we extract a subgraph $G'( V, E', W)$ formed with $V =V' \cup V''$, and arcs $E' \subseteq E \wedge ( (v1, v2) \in E \rightarrow W_{v1, v2} >$ \textit{threshold}$)$. A low weight present a low interest of movement while the most significant weights present the interest of moving from one node to another.

The figure \ref{fig:graph2} presents subgraph of the graph in figure \ref{fig:basicgraph}. Where the bold nodes \textit{Musée de Louvre} and \textit{Tour Eiffel} are the mainstream nodes with their respective support value of $280k$ and $356K$.
The Klosgen value threshold is taken at $0.6$. 

\subsection{Spheres of influence}

In graph theory a sphere of influence of the graph assigns, for each node, a ball centered at that node of neighbors distance $D$ equal to a definite value.
In this study, we create a sphere of influence to all mainstream nodes. A sphere of influence of a mainstream node is an aggregation of its neighboring nodes (mainstream and secondary).

The neighbor distance $D$ of an influence sphere equals to the average length of sequences in the sequences dataset minus one. To process a sphere of influence for a mainstream node $S_mi$, we create an array $A_i$ for each mainstream node $m_i$. The array contains all neighbors of distance $1$ to $D$ from $m_i$.
The table \ref{table:1} shows the result of creating the influence spheres of two mainstream nodes $m_{Tour~Eiffel}$ and  $m_{Musee~du~Louvre}$ the graph figure \ref{fig:graph2}.

\begin{table}[]
\caption{Sphere of Influence of the two mainstream nodes of Figure \ref{fig:graph2}.}
  \label{table:1}
  \centering
\begin{tabular}{|p{4cm}|p{3.5cm}|l|}
\hline
Mainstream node & $m_{Musee~du~Louvre}$     & $m_{Tour~Eiffel}$\\ 
\hline
Neighbors at distance  = $1$            & \begin{tabular}[c]{@{}l@{}} $A_{Musee~du~Louvre}$ = \\ \{Tour Eiffel, \\ Cathédral Notre-Dame, \\ Champs Elysées \}\end{tabular} & \begin{tabular}[c]{@{}l@{}}$A_{Tour~Eiffel}$ =\\ \{Champs Elysées\}\end{tabular}      \\ \hline
Neighbors at distance  = $2$              & none   & \begin{tabular}[c]{@{}l@{}}$A_{Tour~Eiffel}$ =\\ \{Champs Elysées,\\ Cathédral Notre-Dame\}\end{tabular}  \\ \hline
\end{tabular}
\end{table}

\subsection{Similarity measure}

Once all influence spheres are created, we will compare them in pair. Comparing two influence spheres consist to analyze all the nodes composing them and report the percent of similitude into a matrix.

Let $M$ be the matrix of similarity of size $|V'|$, referring to the number of mainstream nodes. The value $M(i,j)$ equal to the percent of nodes in $S_mi$ (i.e. the influence sphere of a mainstream node $m_i$) that are also present in nodes $S_mj$ (i.e. the influence sphere of a mainstream node $m_j$). Thus, the matrix is non-symmetric.
$M(i,j)=1$ means that all the nodes in $S_mi \subseteq S_mj$ and $M(i,j)=0$ means  that $S_mi \cap S_mj = 0$.
Let us take for example the results in table \ref{table:1}. The matrix of similarity $M$ is as follows:
\begin{small}
\[
M =
\begin{blockarray}{cccc}
  &  m_{Musee~du~Louvre} & & m_{Tour~Eiffel}   \\
\begin{block}{l(ccc)}
   m_{Musee~du~Louvre~~} & - &  & 0.66    \\
  m_{Tour~Eiffel}  & 1 &  & -   \\
\end{block}
\end{blockarray}
 \]
 \end{small}
 
\subsection{Profiling}

Based on the similarity matrix, we detect community for purpose to create a tourism profile. The matrix is seen as a graph for the profiling process.
In this study, we are looking for a community detection based on high similarity values and high density of arcs inside communities compared to that of arcs connecting to outside communities.
This distinction between arcs inside the communities and outside is called \textit{modularity}. Then we must optimize the modality in order to have the best possible grouping into communities.
To achieve this, we use the \textit{Louvain clustering} \cite{blondel2008fast} which maximizes modularity for each community.

%
%
%

\section{Experiments}
\label{sec:xp}

We conducted experiments on reviews posted by tourists between $1st$ January $2013$ to $31th$ December $2016$ in  \texttt{Tripadvisor} focused on Paris, French regions.
After cleaning and pre-processing the data, we obtain $1'666'584$ trips of more than $3$ reviews, with an average length of stay of $2.5$ days and an average number of reviews per stays at $4.14$ with close to $40'000$ places of Paris.

\subsection{Measure of interest}

In our study, values of weights of the graph are computed using $Klosgen$ ($Kl$) measure. To compare our results with others methods, we also provide the following measures: $Support$ ($Supp$): depends on the number of occurrences of the rule $X\longrightarrow Y$, $Confidence$ ($Conf$) : is a ratio between the support of the rule on the support of $X$, as a probability of a Markov chain, $Lift$ : provides a value showing how different of a random process the rule is, $Certainty~Factor$ ($CF$): a high Certainty Factor means $Y$ is dependent of $X$ and not another node, $J-Measure~(cross-entropy)$ ($J$): computes the cross-entropy, i.e. the information contain in the rule over all the rules and $Conditional~Entropy$ ($CE$): compute the correlation between $X$ and $Y$ based on the entropy value. 
%

The table \ref{table:2} presents the top $3$ rules based on Support and top $3$ rules based on Klosgen. 
Nodes representing a monuments \textit{Tour Eiffel}, \textit{Musée du Louvre}, \textit{Cathédrale Notre-Dame} and other well-known monuments will be over-represented with Support. Confidence and Lift highest values are a combination of a high support node with one with a very few support.
Same for Certainty Factor, the values between the highest support nodes or node presents in few rules are very low. J-Measure are closed to zero for all rules since rules are numerous for a fixed $X$ or a fixed $Y$. Moreover, the large majority of Conditional Entropy are closed to zero.

As shown, Klosgen measure is not influence by the support of nodes, nor by the number of nodes (entropy tends to zero). Since the value is between $0$ to $1$, Klosgen measure is easily readable and understable. This measure fits perfectly with the notion of \textit{interest}.

\begin{table}[b]
\caption{Measures' results on a sample from the database.}
\label{table:2}
\centering
\begin{tabular}{|l|ccccccc|}
\hline
                &  &  & &   &  &    & \\
\centering Rules           & $Kl$ & $Supp$ & $Conf$ & $Lift$ & $CF$    & $J$    & $CE$
\\\hline
\begin{tabular}[c]{@{}c@{}} Tour~Eiffel $\longrightarrow$  Musée~du~Louvre\end{tabular}         & 0.33       & \textbf{49565}   & \textbf{0.18}  & 0.58         & -0.18 & \textbf{0.03} & \textbf{0.68} \\ 
\hline
\begin{tabular}[c]{@{}c@{}} \small Tour~Eiffel $\longrightarrow$  Cathédrale Notre-Dame\end{tabular}   & 0.57         & \textbf{26129}  & 0.09          & 0.51         & -0.11 & \textbf{ 0.02} & \textbf{0.45} \\ 
\hline
\begin{tabular}[c]{@{}c@{}} Cathédrale Notre-Dame $\longrightarrow$ Tour~Eiffel\end{tabular}    & 0.18         & \textbf{23092} & \textbf{0.20} & 0.45          & -0.43 & \textbf{0.04} & \textbf{0.71} \\ 
\hline
\begin{tabular}[c]{@{}c@{}} Le Bataclan $\longrightarrow$ Place de la République\end{tabular}   & \textbf{0.86}  & 300           & 0.09          & \textbf{2.53} & \textbf{ 0.06}  & 0    & 0.44 \\ 
\hline
\begin{tabular}[c]{@{}c@{}} Hôtel Plaza Athénée  $\longrightarrow$ Hotel George V\end{tabular}  & \textbf{0.85} & 297             & 0.05         & \textbf{4.40} & \textbf{0.04}  & 0    & 0.29 \\ 
\hline
\begin{tabular}[c]{@{}c@{}} Arc de Triomphe $\longrightarrow$ Sacré-Coeur\end{tabular}          & \textbf{0.85} & 1285            & 0.05        & \textbf{1.47} & \textbf{0.02}  & 0             & 0.27 \\ 
\hline
\end{tabular}
\end{table}

\subsection{Mainstream monuments}

To select $K$ mainstream monuments of Paris, we initially order monuments by decreasing order of support and then we compute the cumulative number of reviews per monument of Paris.
The curve displays an elbow which distinguishes on the left side mainstream monuments to secondary monuments. We set the number of mainstream monuments when the elbow start at $K=111$ which represents approximately $73\%$ of all reviews.


\subsection{Sphere of influence}

To create influence spheres, we first start in mainstream monuments then aggregate the neighbors nodes with distance of $D=3$, average length of sequences in the dataset, in our case $4$, minus one.
The similarity matrix and the Louvain clustering are compute on the $111$ corresponding nodes with a Klosgen's threshold  $=0.1$. Since the graph have about $40'000$ nodes, with $111$ mainstream nodes, we do not provide a figure presenting the result. Concerning the similarity matrix, its size is $111*111$, thus we can't show it in this paper.

\subsection{Clustering Analysis }

After applying the Louvain clustering algorithm, we obtain $4$ clusters of similar size, around twenty nodes. We do not regard the $5$ clusters of size less than $3$ and clusters with singleton.
We note that tourist's profiles are bounded to an unique cluster and they do not overlap.

\textit{ Cluster[1]} contains mainly architectural monuments, some can be visited but they are principally known for their aspect like the \textit{Pyramide du Louvre}, \textit{Grand Palais} or \textit{Opéra Garnier}. This cluster contains the main bridges of Paris and typical district like \textit{Montmartre} or \textit{Le Marais}. One monument radically differs is the famous tearoom \textit{Ladurée}. But this restaurant is very close to the monuments of this cluster, which can explain its presence. The profile of this cluster is \textit{Architectural tourism} \cite{specht2014architectural} and \textit{Photography tourism} \cite{picard2016framed}.

 \textit{Cluster[2]} contains architectural monuments and cultural monuments like museums (mostly about history of the country). Some of those monuments also have a religious context like the \textit{Cathédrale Notre-Dame} or \textit{Basilique Sacré-Coeur}. The outsiders are \textit{Tour Eiffel} and \textit{Jardin du Luxembourg}, those monuments are most recent than the others but have a strong cultural aspect. The profile of this cluster is \textit{Heritage tourism} \cite{park2013heritage} and \textit{Religious tourism} \cite{rashid2018religious}.

 \textit{Cluster[3]} contains mostly sport complexes, stadium and cabarets. The outsiders are the restaurant \textit{Le Perchoir}, the train stations \textit{Gare de Lyon} and \textit{Gare de l'Est}. The profile of this cluster is \textit{Sport and Event tourism} \cite{chalip2005sport} and \textit{Recreation tourism}, \textit{Urban tourism} \cite{tribe2020economics}.

 \textit {Cluster[4]} contains the \textit{Grands Magasins de Paris} (Paris department stores), luxury stores and luxury hotels. This cluster clearly refers to shopping and the luxury of Paris, which is a significant attractive aspect of the city. The outsiders are restaurants, but are referred as gastronomic restaurants which refer to luxury. Obviously, this cluster refers to the profiles of \textit{Luxury tourism} and \textit{Culinary tourism} \cite{batat2021role}.

\subsection{Discussions}

\textbf{Impact of Klosgen's threshold}. Since the Klosgen's threshold determines if an arc is kept in the graph, an upper value increase the number of clusters and the number of singleton. When increasing the threshold value, the nodes in a cluster remains together or are split into two or more clusters. The threshold does not modify the profiles but only the number of profiles found.

\textbf{Clustering without computing spheres of influence}. When we compute the Louvain algorithm to the mainstream monuments with Klosgen values without doing the spheres of influence, we obtain $4$ clusters (and some singleton). Some trends can be extract to each cluster, with  several outsiders.
The trends are in order: \textit{Cultural tourism (Heritage)}; \textit{Cultural tourism (Modern Art)} and \textit{Luxury tourism}; \textit{Sport and Event tourism}; \textit{Luxury tourism}. The profiles are less distinct without computing the spheres of influence and some kind of tourist's profiles are present in many clusters.

\textbf{Clustering with other measures}. We also implement the method with support, confidence and lift measures instead of Klosgen measure. The clusters do not have any main trend but are mostly influenced by the metro lines and other public transports of Paris. Clusters reflect also the overtourism in Paris with a cluster with the top $23$ most visited monuments of Paris.

\section{Conclusion}
\label{Conclusion and future works}

When scientists study tourism thanks to social networks data, they used measure based on the frequency to understand the tourists behaviors. We provide a method to extract tourism behaviors using an interest Klosgen measure and by adapting the notion of neighborhood used in social graph to the tourism industry. Our method returns significant results, already observed by tourism management studies in a more limited scale. In future work, we have to enhance our method to refine the profiling, by using a fuzzing clustering to improve the clusters' knowledge discovery. We will additionally use the method to other cities and a vaster area of France to confirm the validity and efficiency of the proposed method.

%
%
%
\bibliographystyle{splncs04}
\bibliography{WISE}
\end{document}